\documentclass[]{spie}  
\usepackage[]{graphicx}

\title{The Speedster-EXD - A New Event-Triggered Hybrid CMOS X-ray Detector} 

\author{Christopher V. Griffith\supit{a}, Abraham D. Falcone\supit{a}, Zachary R. Prieskorn\supit{a}, David N. Burrows\supit{a}
\skiplinehalf
\supit{a}The Pennsylvania State University, 525 Davey Lab, University Park, PA, USA; \\
}

\authorinfo{Further author information: (Send correspondence to C.V.G.)\\C.V.G.: E-mail: cvg5124@psu.edu\\ }

\begin{document} 
  \maketitle 

\begin{abstract}
We present preliminary characterization of the Speedster-EXD, a new event driven hybrid CMOS detector (HCD) developed in collaboration with Penn State University and Teledyne Imaging Systems.  HCDs have advantages over CCDs including lower susceptibility to radiation damage, lower power consumption, and faster read-out time to avoid pile-up. They are deeply depleted and able to detect x-rays down to approximately 0.1 keV. The Speedster-EXD has additional in-pixel features compared to previously published HCDs including: (1) an in-pixel comparator that enables read out of only the pixels with signal from an x-ray event, (2) four different gain modes to optimize either full well capacity or energy resolution, (3) in-pixel CDS subtraction to reduce read noise, and (4) a low-noise, high-gain CTIA amplifier to eliminate interpixel capacitance crosstalk.  When using the comparator feature, the user can set a comparator threshold and only pixels above the threshold will be read out.  This feature can be run in two modes including single pixel readout in which only pixels above the threshold are read out and 3x3 readout where a 3x3 region centered on the central pixel of the X-ray event is read out.  The comparator feature of the Speedster-EXD increases the detector array effective frame rate by orders of magnitude.   The new features of the Speedster-EXD hybrid CMOS x-ray detector are particularly relevant to future high throughput x-ray missions requiring large-format silicon imagers.

\end{abstract}


\section{INTRODUCTION}
\label{sec:intro} 
Future x-ray missions, such as SMART-X\cite{2012SPIE.8443E..16V}, will have much larger collecting areas than current missions such as Chandra and XMM-Newton.  Future missions will focus on observing much fainter objects, such as new x-ray bursts at early times in the universe (z $>$ 7) which will explore cosmic structure and the birth of stars.  In addition to the faint objects, future x-ray missions will observe bright objects such as blazars.  With the increased collecting area of future missions combined with the observation of bright objects such as blazars, current x-ray CCDs do not have fast enough read out times to keep up with the amount of x-rays that will be collected.  

	Hybrid CMOS x-ray detectors offer the fast read out times needed for future high throughput x-ray space missions.  Their pixel architecture enables the speed to capture the source image before multiple photons saturate a pixel.  Hybrid CMOS detectors also require less power and are more radiation hard than CCDs, adding to mission lifetimes. A more thorough overview of the advantages of hybrid CMOS x-ray detectors can be seen in Falcone et al. 2014\cite{2014abe}.
	
	In this paper, we present two Speedster-EXD hybrid CMOS detectors and their characterization. The Speedster-EXD detector has new in-pixel circuitry that includes a CTIA amplifier to eliminate interpixel capacitance crosstalk, in-pixel CDS subtraction to reduce noise, four different gain modes, and an in-pixel comparator that enables the read out of only pixels with signal from an x-ray event.  The detector can be run in full frame read out mode where all pixels are read out, and in sparse mode where only pixels which contain an x-ray event are read out.  We discuss the performance of each of these features and present the measured read noise, energy resolution, interpixel capacitance, and gain variation. 
	
\section{CHARACTERIZING A HYBRID CMOS X-RAY DETECTOR FOR X-RAY ASTRONOMY}
To demonstrate the capabilities of the Speedster-EXD HCDs, we tested four key parameters of each detector.

\subsection{Energy Resolution}
Energy resolution is one of the most important characteristics of an x-ray detector.  CCDs are currently at near Fano limited resolution ($\Delta$E/E$\sim$2.0\% at 5.9 keV) so the HCD detectors must approach this resolution to compete with CCDs.  Previous generation Si HCDs were affected by interpixel capacitance (IPC) and high read noise which degraded their energy resolution. 

\subsection{Interpixel Capacitance (IPC)}
Interpixel Capacitance(IPC) is caused by unintended, parasitic capacitances between adjacent pixels inside the silicon absorber array\cite{2006SPIE.6276E..13F}.  The interpixel capacitance of our detector can affect the PSF and can make single pixel x-ray events appear as poor cross events.   This signal spread causes a degradation in energy resolution as more pixels must be read out so it is essential detectors have low IPC.  


\subsection{Read Noise}
Read noise arises from the charge fluctuations at the amplifier of our detector. These fluctuations arise at the time interval between when the charge is read out by the ROIC and when it is read into a computer as an A/D channel number.  The read noise measurement sets the noise floor and is a factor in detector energy resolution.  CCDs currently have lower read noise at slow read out speeds so it is important HCDs continue to lower their read noise. 

\subsection{Gain Variation}
Gain variation between pixels causes degradation of energy resolution.  Gain variation appears to be a low, but non-negligible, effect in hybrid CMOS detectors.  Unlike CCDs, hybrid CMOS have an amplifier is every pixel which can vary and cause small fluctuations between pixels. These small fluctuations can cause differences in energy measurements of x-ray photons depending on which pixel they are detected in.  By measuring the gain variation, we can determine how much it is degrading the energy resolution on HCDs.

\section{SPEEDSTER-EXD HYBRID CMOS X-RAY DETECTOR}
\begin{figure}[h!]
\vspace{1cm}
\center
\includegraphics[width=8cm,height=6cm,scale=.75]{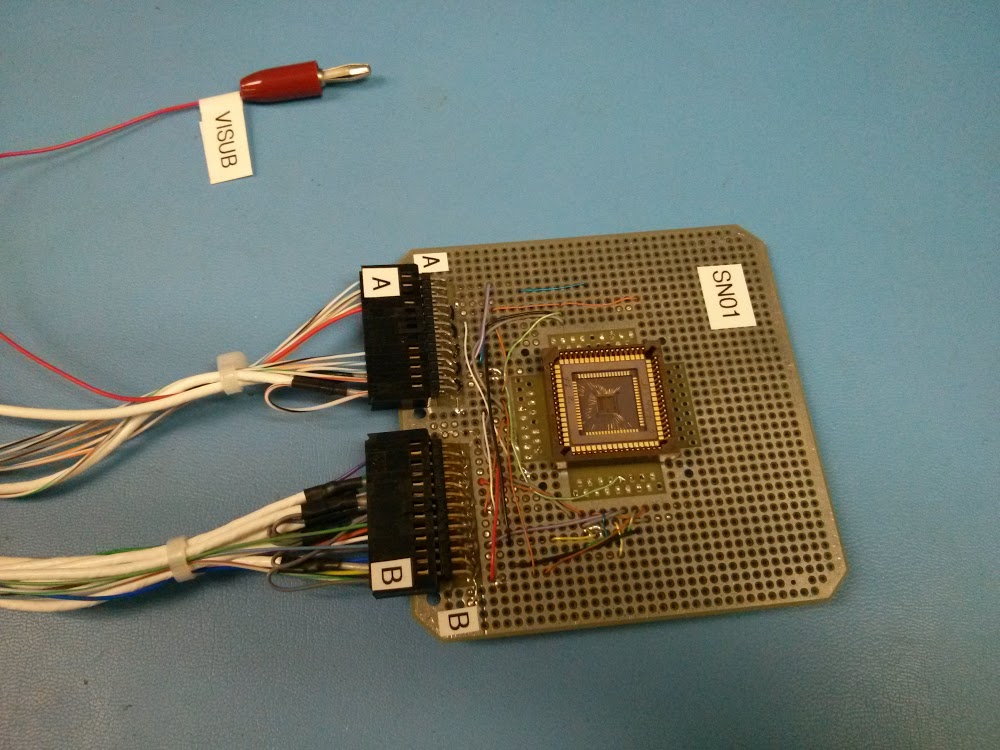}
\caption{Speedster-EXD hybrid CMOS detector located in the dewar breadboard.}
\label{speedster_board}
\end{figure}

\begin{figure}[h!]
\vspace{1cm}
\center
\includegraphics[width=12cm,height=8cm,scale=.75]{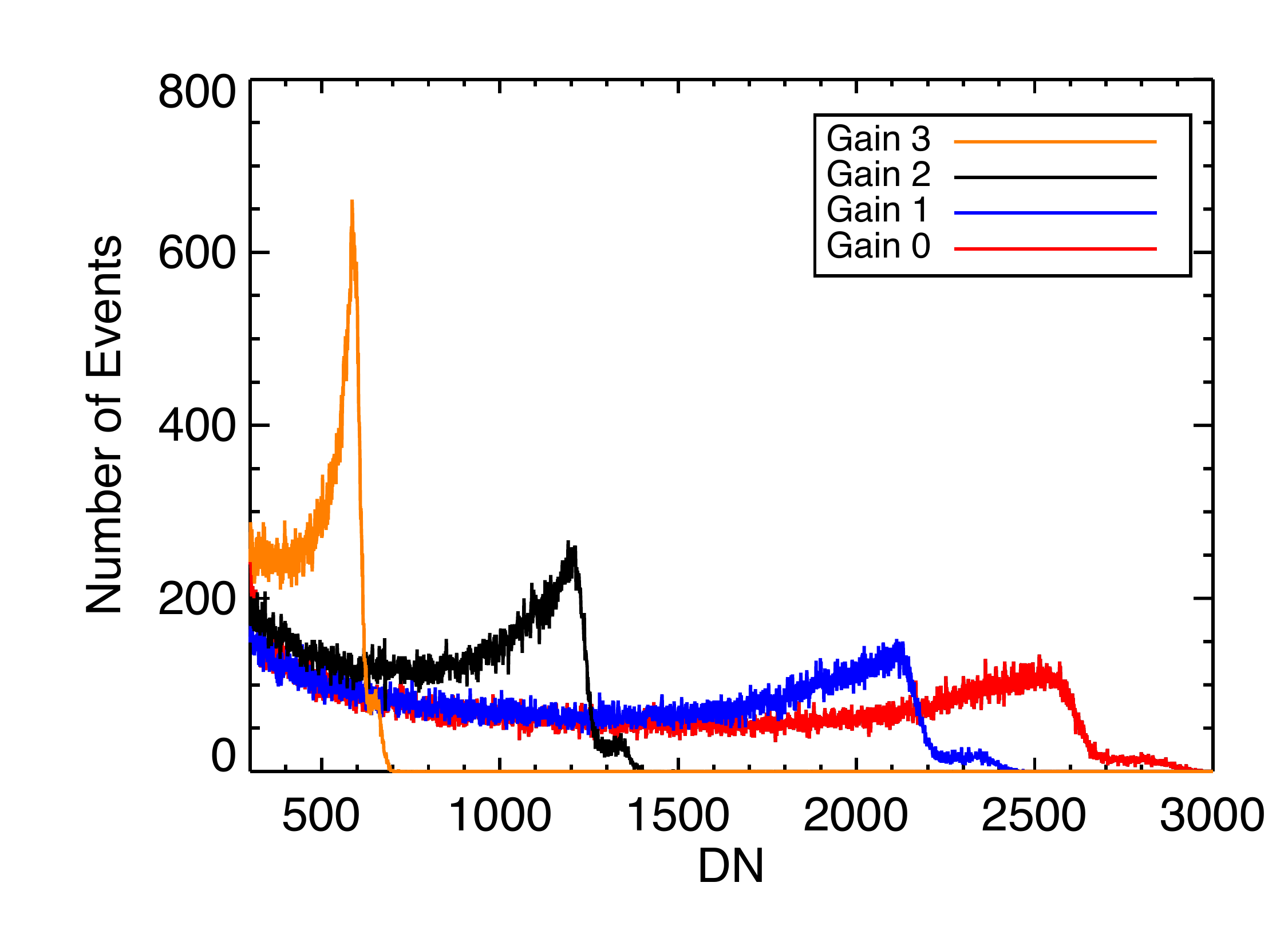}
\caption{Raw unfiltered spectra from all four gain modes in the FPA 17017 Speedster-EXD detector.  The Mn K$\alpha$ peaks range from 550DN to 2500DN.  }
\label{gain_modes}
\end{figure}

\begin{figure}[htbp]
\begin{center}
  \begin{minipage}[b]{0.45\linewidth}
    \centering
    \includegraphics[width=\linewidth]{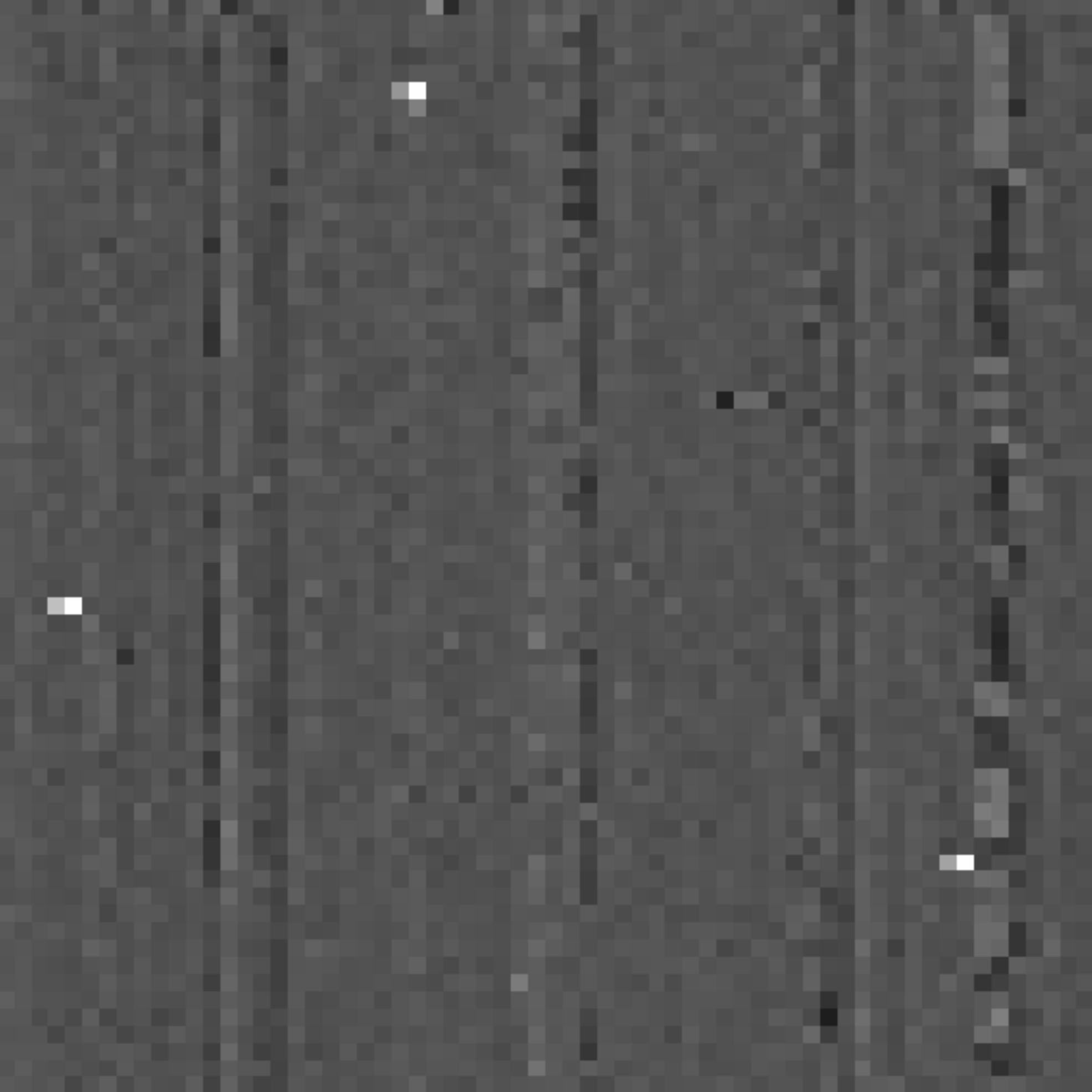}
    \caption{Full Frame Read Out Mode: The comparator is set below the read noise floor so all pixels on the detector are read out.  You can see three $^{55}$Fe x-ray events along with the other pixels where no x-ray signal was detected.  }
    \label{full_frame_fig}
  \end{minipage}
  \hspace{0.5cm}
  \begin{minipage}[b]{0.45\linewidth}
    \centering
    \includegraphics[width=\linewidth]{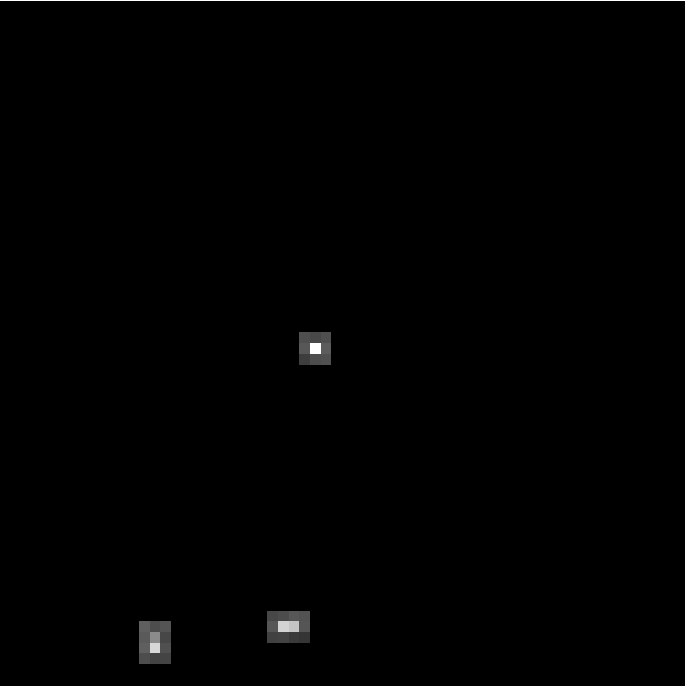}
    \caption{Sparse 3x3 Read Out Mode: The comparator is set above the noise floor so only x-ray events are read out.  The above image was taken in 3x3 sparse mode so only three $^{55}$Fe x-ray events are read out along with the 3x3 region around them.}
    \label{sparse_fig}
  \end{minipage}
  \end{center}
\end{figure}

The Speedster-EXD hybrid CMOS detector is a 64 x 64 40$\mu$m pitch prototype detector fabricated by Teledyne Imaging Systems.  A picture of the Speedster-EXD detector located in its dewar breadboard can be seen in Figure \ref{speedster_board}. The detector has many features that improve upon our previous HCDs.  These features include:
\begin{itemize}

\item {\bf In-Pixel Correlated Double Sampling (CDS) Subtraction:} The circuitry in each pixel on the Speedster-EXD includes CDS subtraction in order to reduce the reset noise of the detector and thereby decrease the overall noise of the detector. The circuitry allows for the measurement of the shifting voltage level that results from the noise associated with the reset of the pixel. This measured noise is then subtracted off the integrated charge in each pixel. 

\item {\bf CTIA Amplifier:} The Speedster-EXD uses a high-gain, low-noise CTIA amplifier.  This amplifier is used to eliminate signal spreading between pixels known as Interpixel Capacitance Crosstalk (IPC).  We have seen IPC degrade the energy resolution on our previous generation of detectors\cite{2013NIMPA.717...83P, 2012SPIE.8453E..0FG} and this effect is significantly reduced on the Speedster-EXD detector (See Section \ref{ipc}).  Unlike the source follower on our previous generation HCDs, the CTIA amplifier is held at a constant voltage during integration time.  This constant voltage on the input node of the CTIA amplifier prevents signal spreading between pixels.  

\item {\bf Four Gain Modes:} In-pixel circuitry allows the user to select four different gain modes ranging from $\sim$2.7 e$^{-}$/DN to $\sim$0.6 e$^{-}$/DN.  This allows the user to optimize for full well capacity or energy resolution.  A low gain mode allows the collection of more x-ray photons before filling up the full-well since the e$^{-}$/DN ratio is high, and a high gain mode offers better energy resolution since the x-ray energy will peak at a higher DN level.  A plot of the raw spectra of Mn K$\alpha$ and K$\beta$ lines from all four gain modes can be seen in Figure \ref{gain_modes}.

\item {\bf Sparse Read Out:} One of the key features of the Speedster-EXD is sparse readout, the ability to read out only pixels which contain an x-ray
event. The user defines a threshold and only pixels which exceed this threshold are read out. Sparse readout can be
run in two modes including single pixel readout, where only the pixels above the set threshold are read out and 3x3
readout, where the 3x3 region centered on the central pixel of the x-ray event are read out.  The sparse feature is controlled using an in-pixel comparator which is autozeroed to a low energy cutoff of 200 eV to avoid any non uniformity between comparators. Two example images with sparse mode turned on and off can be seen in Figures \ref{full_frame_fig} and \ref{sparse_fig}.
 
\end{itemize}

Each Speedster-EXD detector has an aluminum optical blocking filter deposited directly on the silicon substrate.  The Speedster-EXD has up to a 10kHz frame rate when running in full frame read out mode.  The effective frame rate is faster when the sparse mode is used to read out only pixels with x-ray events.


\section{EXPERIMENT SETUP}
\begin{figure}[h!]
\vspace{1cm}
\center
\includegraphics[width=8cm,height=6cm,scale=.75]{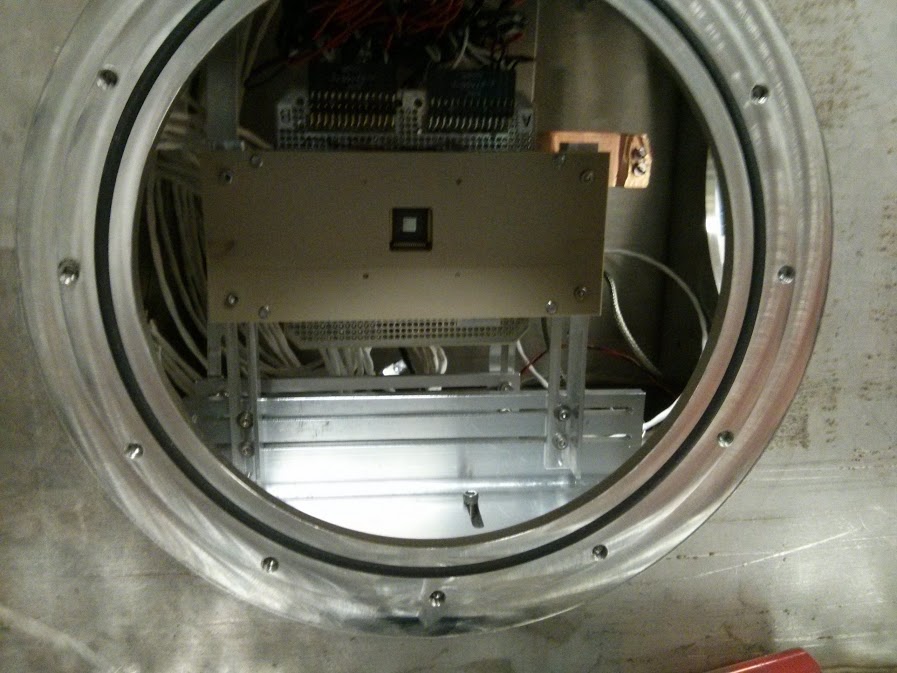}
\caption{Speedster-EXD hybrid CMOS detector located in the ``cube" test stand.}
\label{speedster_cube}
\end{figure}

We test two Speedster-EXD detectors referred by their serial number; FPA 17014 and FPA 17017.
We modified our existing ``cube" test stand at Penn State University in order to cool the detectors and expose them to x-rays.  Figure \ref{speedster_cube} shows a Speedster-EXD mounted in the ``cube" chamber.  Before we take exposures, we pump the chamber down to below 10$^{-6}$ Torr and cool the detector to 150K using liquid nitrogen.  We then expose the detectors to x-rays using an $^{55}$Fe radiation source (which produces Mn K$\alpha$ and K$\beta$ x-rays).   We run both detectors at a bias voltage of 15V.  The Speedster-EXD detectors are controlled using a TIS developed camera electronic (Detector Interface Block, DIB) and specially designed software.  Using the software and DIB, the user can send registers which control the mode of operation (sparse mode or full frame read out) and also settings such as the threshold above which pixels will be read out in sparse mode, exposure time, etc.  In sparse mode, the user can also determine whether to run single pixel read out mode, where only the pixels which have signal above the set threshold are read out or 3x3 mode, where the eight surrounding pixels are read out along with the pixel containing the x-ray signal. 

\begin{figure}[h!]
\vspace{1cm}
\center
\includegraphics[width=8cm,height=10cm,scale=.75]{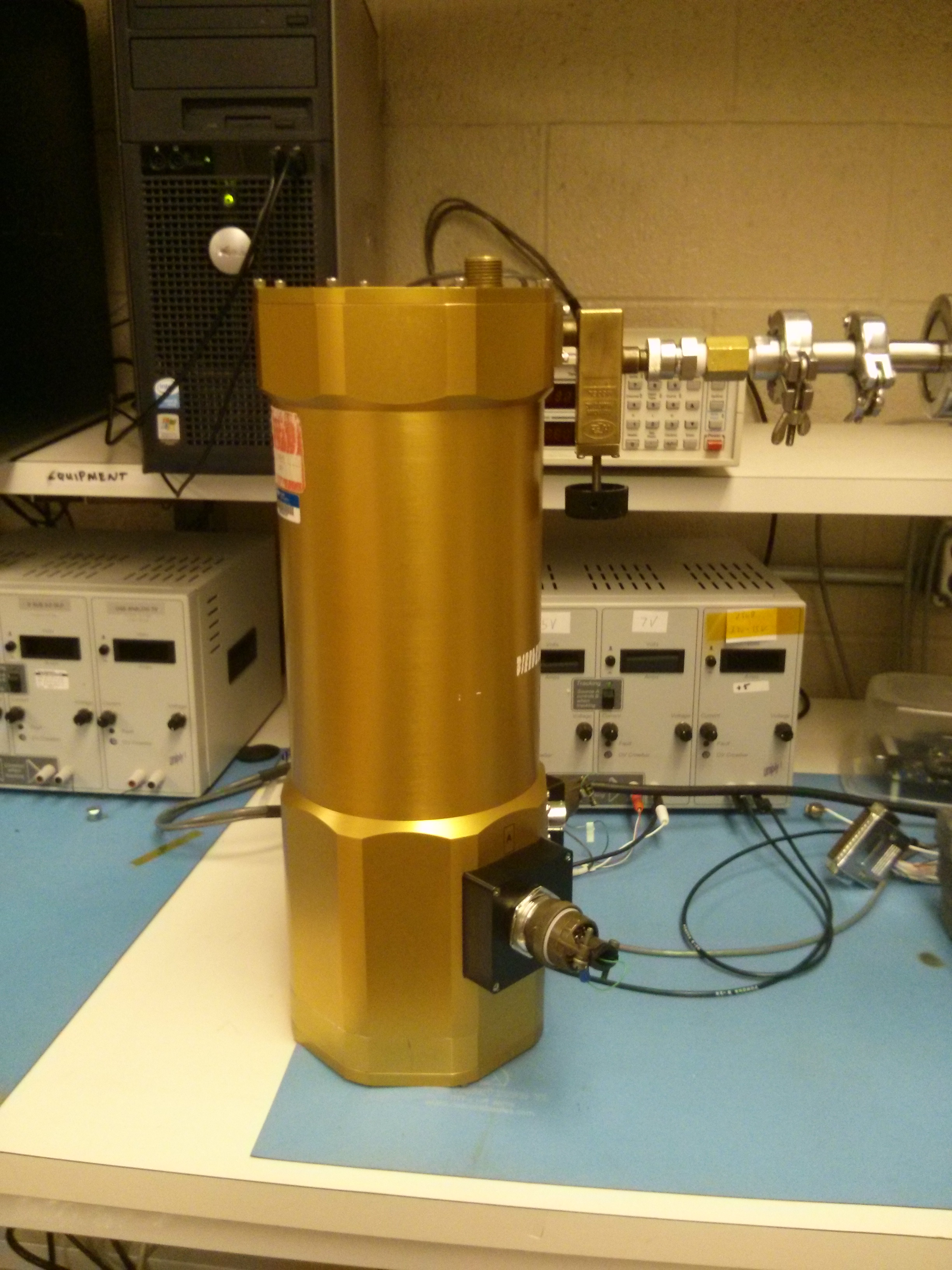}
\caption{IR Labs HDL5 dewar used for Speedster-EXD characterization. }
\label{teledyne_dewar}
\end{figure}

We also ran tests using a IR Labs HDL-5 dewar while at Teledyne.  This dewar cooled the detector for much longer periods of time than the ``cube" chamber, which allowed for longer data taking runs.  The HDL-5 dewar is pictured in Figure \ref{teledyne_dewar}.  

\section{DATA REDUCTION}
The Speedster-EXD detector is operated in two modes.  The first mode is full frame read out mode where the comparator is set below the noise floor (see Figure \ref{full_frame_fig}).  This setting allows all of the pixels to be read out.  The second mode is sparse mode where the comparator is set above the noise floor allowing only pixels which have detected x-rays to be read out (see Figure \ref{sparse_fig}).  We present data in sparse 3x3 read out mode so we can correctly characterize the events.  Both full frame read out mode and sparse mode require their own data reduction techniques.

\subsection{Full Frame Read Out Mode}
\label{full_frame_mode}
In full frame read out mode, we construct a bias image from the pixels in each image which have no x-ray signal in them and average them together.  We then subtract this bias image from each image.  Once we have the bias subtracted image, we set a primary threshold for an x-ray event to 5$\sigma$, where $\sigma$ is the read noise, and set the criteria that the pixel containing x-ray signal must be a local maxima.  We then set a secondary threshold of 3$\sigma$, where if any of the eight pixels surrounding the primary pixel are measured to be above this threshold those pixels' signal are included in the total event signal.   We then ``grade" the events depending on how many of the eight surrounding pixels are included in the event using the Swift XRT grading scheme\cite{2005SSRv..120..165B}.  In this paper, we present single pixel events (grade 0 events), where x-ray charge from the event remains in the center pixel and singly split pixel events (grade 1-4 events) where the x-ray charge diffuses between the center pixel and one of the four adjoining pixels. 

\subsection{Sparse Mode} \label{sparse_mode}
In sparse mode, we could not construct a bias image from the pixels which contain no x-ray signal since those pixels are not read out.  We instead set the comparator to below the noise floor (full frame read out mode) and captured bias images.  Then we moved the comparator above the noise floor to read out only x-ray events.  We average the bias images captured in full frame read out mode and subtract the averaged bias image from the sparse images.  We used the same 5$\sigma$ primary threshold and 3$\sigma$ secondary threshold to properly grade the events as explained in Section \ref{full_frame_mode}.

\section{ANALYSIS}

\subsection{Read Noise}
We found the read noise by taking dark exposures with the detector. We cooled the detector to 150K, put it under vacuum, and took 1000 exposures with no x-ray source.  We then took the standard deviation of each pixel over the 1000 exposures and histogrammed the result.  We found the mean of the histogram to measure the read noise in Digital Number (DN).  The electron/DN conversion factor was found using the fact that 1616e$^{-}$ are expected from each Mn K$\alpha$ photon.  We found the mean of the Mn K$\alpha$ line for each detector at each gain setting and used the conversion factor to convert read noise to electrons.  The measured read noise for two Speedster detectors in all four gain modes are shown in Table \ref{read_noise}.


\begin{table}[ht]
\begin{minipage}[b]{0.5\linewidth}\centering
\begin{tabular}{c|c}
\multicolumn{2}{c}{{\bf FPA 17014}}\\
\hline
Gain Mode & Read Noise \\
\hline
Gain 0 & 15.2 e$^{-}$\\
Gain 1 & 14.2 e$^{-}$\\
Gain 2 &13.0 e$^{-}$ \\
Gain 3 & 16.3 e$^{-}$\\

\end{tabular}
\end{minipage}
\hspace{0.1cm}
\begin{minipage}[b]{0.0\linewidth}
\centering
\begin{tabular}{c|c}
\multicolumn{2}{c}{{\bf FPA 17017}}\\
\hline
Gain Mode & Read Noise \\
\hline
Gain 0 & 12.3 e$^{-}$\\
Gain 1 & 11.2 e$^{-}$\\
Gain 2 &13.1 e$^{-}$ \\
Gain 3 & 16.9 e$^{-}$\\
\end{tabular}
\end{minipage}
\caption{Measured read noise for two Speedster-EXD detectors in full frame read out mode.  }
\label{read_noise}
\end{table}

\subsection{Energy Resolution}
\subsubsection{Full Frame Read Out Mode}
Before constructing histograms of the Mn K$\alpha$ and K$\beta$ peaks, we first had to grade the events that were detected.  As mentioned in Section \ref{full_frame_mode}, we set a primary and secondary threshold for including charge in the x-ray events and then graded the events.  We use grade 0 through 4  events (single pixel and singly split events) in our spectra. 

We measured the energy resolution by first fitting multiple Gaussians to the Mn K$\alpha$ and K$\beta$ peaks and calculating the Full Width Half Max (FHWM) = 2$\sqrt{\rm2ln2}\sigma$ where sigma is the standard deviation of the fitted Gaussian of the energy peaks. We then took the FWHM and divided it by the energy at the peak of the Gaussian to get a value for $\Delta$E/E to get the percent values quoted in this paper.   The measured energy resolution for the two Speedster-EXD detectors in the highest gain mode (gain 0) and the energy resolution for FPA 17014 in the second highest gain mode (gain 1) can be seen in Tables \ref{energy_res_014} and \ref{energy_res_017}.  These measurements were taken in full frame read out mode and use only grade 0 events (single pixel events) and grade 0-4 events (single pixel events and singly split events).  

\begin{figure}[ht!]
\vspace{1cm}
\center
\includegraphics[width=12cm,height=8cm,scale=.75]{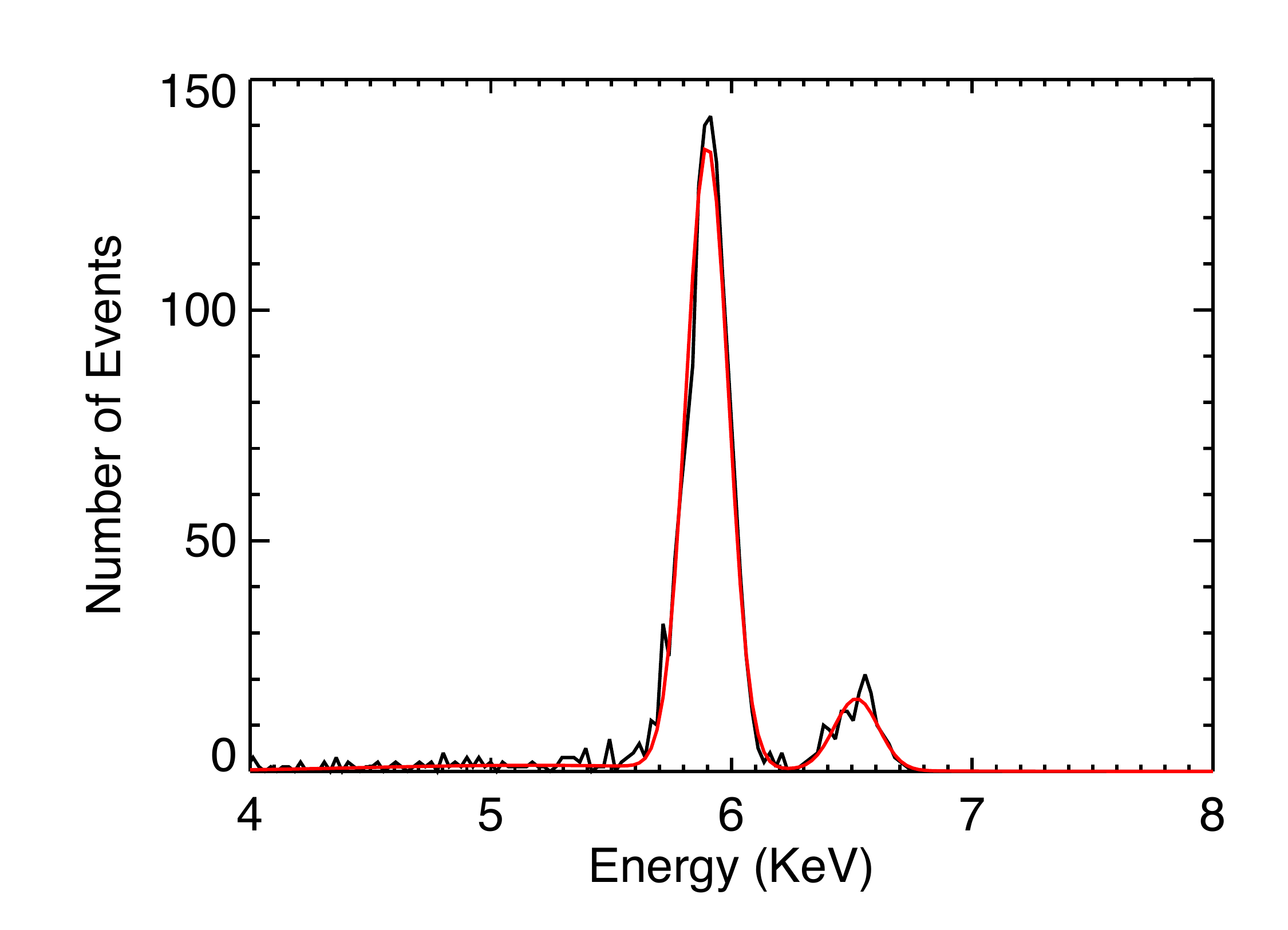}
\caption{Manganese K$\alpha$ and K$\beta$ spectra taken in gain 1 in full frame read out mode using only grade 0 events on FPA 17014.  The measured $\Delta$E/E = 3.5\% with a FWHM = 206 eV.  }
\label{gain1_spec}
\end{figure}

We measure the best energy resolution in full frame read out mode to be $\Delta$E/E = 3.5\% with a FWHM = 206 eV shown in Figure \ref{gain1_spec} using only grade 0 events in the second highest gain mode (gain 1) on FPA 17014.



\begin{table}[ht]
\begin{center}
\begin{tabular}{|c|c|c|}
\multicolumn{3}{c}{{\bf FPA 17014 (Full Frame Read Out)}}\\

\hline
Gain Mode &FWHM (eV) & $\Delta$E/E\\

\hline
Gain 0 (Grade 0 Events) & 215eV & 3.7\% \\
Gain 1 (Grade 0 Events) & 206eV & 3.5\% \\
\hline
Gain 0 (Grade 0-4 Events) & 292eV & 5.0\% \\
Gain 1 (Grade 0-4 Events) & 276eV & 4.7\% \\
\hline

 \end{tabular}
  \vspace{.1cm}
 \caption{Measured energy resolution for the FPA 17014 Speedster-EXD detector in the highest gain mode (gain 0) and the second highest gain mode (gain 1) in full frame read out mode using grade 0 events (single pixel events) and grade 0-4 events (single pixel events and singly split events).}
\label{energy_res_014}
\end{center}
\end{table}

\begin{table}[ht]
\begin{center}
\begin{tabular}{|c|c|c|}
\multicolumn{3}{c}{{\bf FPA 17017 (Full Frame Read Out)}}\\
\hline
Gain Mode & FWHM (eV) & $\Delta$E/E\\
\hline
Gain 0 (Grade 0 Events) & 324 eV & 5.5\% \\
\hline
Gain 0 (Grade 0-4 Events)& 427eV & 7.3\% \\
\hline
\end{tabular}
 \vspace{.1cm}
\caption{Measured energy resolution for the FPA 17017 Speedster-EXD detector in the highest gain mode (gain 0) and the second highest gain mode (gain 1) in full frame read out mode using grade 0 events (single pixel events) and grade 0-4 events (single pixel events and singly split events)}
\label{energy_res_017}
\end{center}

\end{table}







\subsubsection{Sparse Mode}
We next looked at the energy resolution in sparse 3x3 mode.  As explained in Section \ref{sparse_mode}, we had to use a constructed bias image from full frame read out mode in order to process the images correctly.  Once the images were bias subtracted, we used the same grading thresholds as we used in full frame read out mode.  We measure the best energy resolution to be $\Delta$E/E = 4.2\% with a FWHM = 220 eV in sparse read out mode in the highest gain mode (gain 0) on FPA 17014 using grade 0 events.  


\subsection{Interpixel Capacitance}
\label{ipc}
We measure the Interpixel Capacitance Crosstalk (IPC) using the ``paired pixel" method described in Prieskorn et al 2013\cite{2013NIMPA.717...83P}.  This method measures the standard deviation of up and down pixels and the left and right pixels in a 3x3 event. If the standard deviation of the up and down pixels and the standard deviation of the left and right pixels were less than 1$\sigma$ of the read noise, we classified the event as one which was suitable for measuring IPC.  We used the ``paired pixel" method to ensure we were not including ``split" events (grade 1-4 events) which occur when an x-ray charge is split between two or more pixels due to the position on the pixel where the x-ray is detected.  We wanted to use only single pixel events which were solely affected by IPC.  We measured the IPC using FPA17014 in the highest gain setting (gain 0).  We averaged and normalized all of the IPC events to get the kernel seen in Table \ref{table_ipc}.  
\begin{table}[ht!]

\begin{center}
\begin{tabular}{|c|c|c|}

\multicolumn{3}{c}{{\bf Speedster-EXD FPA 17014 IPC Measurement}}\\
\hline
0.0002 $\pm$0.010&-0.0007 $\pm$ -0.001& -0.0001 $\pm$ -0.010 \\
\hline
0.002 $\pm$ 0.009&1.0045 $\pm$ 0.13&0.002 $\pm$ 0.009\\
\hline
0.0002 $\pm$ 0.009&-0.008 $\pm$ -0.013& -0.0002 $\pm$ -0.009\\
\hline

\end{tabular}
\vspace{.2cm}
\caption{Charge distribution of x-ray events used for IPC measurement for FPA 17014 in gain 0 (highest gain) mode.  The events that we deemed suitable for measuring IPC were all averaged and then the kernel was normalized.  We see that nearly all of the charge is retained in the center pixel and the IPC is $<$1\%.  The negative values seen are an artifact of our bias image subtraction. }
\label{table_ipc}
\end{center}

\end{table}

We find that nearly all of the charge is retained in the center pixel and the interpixel capacitance is {\bf $<$1\%} in the neighboring pixels.  This result is a big improvement over our previous 18$\mu$m pixel pitch H1RGs,  which showed 4-9\% signal spreading, and our 36$\mu$m pitch H2RG, which showed 1.7-1.8\% signal spreading.\cite{2013NIMPA.717...83P, 2012SPIE.8453E..0FG} 

\subsection{Gain Variation} 
We next wanted to calculate the gain variation of the Speedster-EXD 17014 to see if it was a possible contribution to $\Delta$E/E.  We took around 1,125,000 images using the FPA 17014 in full frame read out mode using the gain 3 setting (lowest gain setting) in order to get $\sim$30 single pixel (grade 0) Mn K$\alpha$ x-ray events in each pixel.  Once we had taken all of the images, we then histogrammed all of the singe pixel events in each pixel and fit a Gaussian to calculate the mean.  However, even with the high number of images, we still did not have enough single pixel events to provide a good Gaussian fit.  We then changed the criteria to add singly split events (grade 1-4).  If a pixel was measured between 200DN and 1000DN (Mn K$\alpha$ peak is at 650DN in gain 3 mode) and its neighboring pixels were less than 15DN ($\sim$1.5$\sigma$) then it was counted as a single pixel event and only the charge in the center pixel was used.  We also added the criterion that if a pixel was between 200DN and 1000DN and its neighboring pixels were less than 40DN ($\sim$4$\sigma$) then the neighboring pixels between 15DN and 40DN were added to the center pixel and used in the gain variation calculation.  Once we had enough events in each pixel, we then fit two Gaussians to the histogram of each pixel, one for the Mn K$\alpha$ peak and one for the Mn K$\beta$ peak.  We tied the parameters of the K$\beta$ peak to the K$\alpha$ peak to reduce the number of free parameters.  We used the mean of the K$\alpha$ Gaussian fit to produce a gain variation map.  We found the standard deviation of the gain variation map and calculated the gain variation to be {\bf 2.6\% $\pm$ 2\%}.

We plan to use the IR Labs HDL-5 dewar pictured in Figure \ref{teledyne_dewar} to calculate the gain variation further.  We can run the speedster cold for longer using this dewar so we can get much better statistics for the gain variation calculation for both FPA 17014 and FPA 17017.

\section{Conclusions}  
We find the Speedster-EXD hybrid CMOS x-ray detector and its sparse circuitry to be fully functional.  We measure the best energy resolution to date on a hybrid CMOS x-ray detector: $\Delta$E/E = 3.5\% at 5.9keV in full frame read out mode.  We also find the sparse circuitry to have comparable energy resolution of $\Delta$E/E = 4.2\% in 3x3 sparse read out mode.  We measure the interpixel capacitance to be $<$1\% which is the lowest reported IPC on a Si hybrid CMOS detector so we can now better characterize x-ray events.  We plan on fully characterizing both Speedster-EXD detectors in the coming months, which will include energy resolution measurements at all gain modes, dark current measurements, further gain variation measurements, and further characterization of the sparse read out feature.   Overall the Speedster-EXD detector displays the potential of hybrid CMOS detectors.  The in-pixel circuitry including CDS subtraction, four gain modes, and sparse read out shows the capability of hybrid CMOS detectors.  X-ray events can now be correctly characterized with the low IPC present in this new generation of detectors and we continue to improve the energy resolution.  Coupled with their fast read out, radiation hardness, and low power, hybrid CMOS detectors are a viable option for many future high-throughput x-ray space missions.    

\section{Acknowledgements}
We gratefully acknowledge Teledyne Imaging Systems especially Vincent Douence for providing useful insight and assistance with the Speedster-EXD detector.  This work is supported by NASA grants NNX08AI64G and NNX11AF98G.

\bibliography{speedster_SPIE_paper}   

\begin{thebibliography}{1}

\bibitem{2012SPIE.8443E..16V}
A.~{Vikhlinin}, P.~{Reid}, H.~{Tananbaum}, D.~A. {Schwartz}, W.~R. {Forman},
  C.~{Jones}, J.~{Bookbinder}, V.~{Cotroneo}, S.~{Trolier-McKinstry},
  D.~{Burrows}, M.~W. {Bautz}, R.~{Heilmann}, J.~{Davis}, S.~R. {Bandler},
  M.~C. {Weisskopf}, and S.~S. {Murray}, ``{SMART-X: Square Meter Arcsecond
  Resolution x-ray Telescope},'' in {\em Society of Photo-Optical
  Instrumentation Engineers (SPIE) Conference Series},  {\em Society of
  Photo-Optical Instrumentation Engineers (SPIE) Conference Series} {\bf 8443},
  Sept. 2012.

\bibitem{2014abe}
A.~{Falcone et al. 2014 (these proceedings)}

\bibitem{2006SPIE.6276E..13F}
G.~{Finger}, R.~{Dorn}, M.~{Meyer}, L.~{Mehrgan}, A.~F.~M. {Moorwood}, and
  J.~{Stegmeier}, ``{Interpixel capacitance in large format CMOS hybrid
  arrays},'' in {\em Society of Photo-Optical Instrumentation Engineers (SPIE)
  Conference Series},  {\em Society of Photo-Optical Instrumentation Engineers
  (SPIE) Conference Series} {\bf 6276}, July 2006.

\bibitem{2013NIMPA.717...83P}
Z.~{Prieskorn}, C.~V. {Griffith}, S.~D. {Bongiorno}, A.~D. {Falcone}, and D.~N.
  {Burrows}, ``{Characterization of Si hybrid CMOS detectors for use in the
  soft X-ray band},'' {\em Nuclear Instruments and Methods in Physics Research
  A}~{\bf 717}, pp.~83--93, July 2013.

\bibitem{2012SPIE.8453E..0FG}
C.~V. {Griffith}, S.~D. {Bongiorno}, D.~N. {Burrows}, A.~D. {Falcone}, and
  Z.~R. {Prieskorn}, ``{Characterization of an x-ray hybrid CMOS detector with
  low interpixel capacitive crosstalk},'' in {\em Society of Photo-Optical
  Instrumentation Engineers (SPIE) Conference Series},  {\em Society of
  Photo-Optical Instrumentation Engineers (SPIE) Conference Series} {\bf 8453},
  July 2012.

\bibitem{2005SSRv..120..165B}
D.~N. {Burrows}, J.~E. {Hill}, J.~A. {Nousek}, J.~A. {Kennea}, A.~{Wells},
  J.~P. {Osborne}, A.~F. {Abbey}, A.~{Beardmore}, K.~{Mukerjee}, A.~D.~T.
  {Short}, G.~{Chincarini}, S.~{Campana}, O.~{Citterio}, A.~{Moretti},
  C.~{Pagani}, G.~{Tagliaferri}, P.~{Giommi}, M.~{Capalbi}, F.~{Tamburelli},
  L.~{Angelini}, G.~{Cusumano}, H.~W. {Br{\"a}uninger}, W.~{Burkert}, and G.~D.
  {Hartner}, ``{The Swift X-Ray Telescope},'' {\em Space Science Reviews}~{\bf
  120}, pp.~165--195, Oct. 2005.

\end{thebibliography}
\bibliographystyle{spiebib}   

\end{document}